\documentclass[11pt]{article}

\usepackage{graphicx}
\usepackage{bm}

\begin{document}


\title{Edge states of Sr$_2$RuO$_4$ detected by in-plane tunneling spectroscopy}

\author{
\textsc{Satoshi. Kashiwaya},$^{1}$ 
\textsc{Hiromi Kashiwaya},$^{1}$ 
\textsc{Hiroshi Kambara},$^{2}$\\
\textsc{Tsuyoshi Furuta},$^{3}$ 
\textsc{Hiroshi Yaguchi},$^{3}$ 
\textsc{Yukio Tanaka},$^{4}$ \\
and \textsc{Yoshiteru Maeno}$^{5}$}
\maketitle


$^1$National Institute of Advanced Industrial Science and Technology (AIST), Tsukuba 305-8568, Japan\\
$^2$Faculty of Education, Shinshu University, Nagano 380-8544, Japan\\
$^3$Department of Physics, Faculty of Science and Technology, Tokyo University of Science,
Noda 278-8510, Japan\\
$^4$Department of Applied Physics, Nagoya University, Nagoya 464-8603, Japan\\
$^5$Department of Physics, Kyoto University, Kyoto 606-8502, Japan


\date{\today}

\begin{abstract}
We perform tunneling spectroscopy of Sr$_2$RuO$_4$ searching for the edge states peculiar to topological superconductivity.
Conductance spectra obtained on Sr$_2$RuO$_4$/Au junctions fabricated using $in$ $situ$ process show broad humps indicating the successful detection of $a$-axis edge of 1.5K phase.
Three types of peak shape are detected: domelike peak, split peak and two-step peak.
By comparing the experiments with predictions for unconventional superconductivity, these varieties are shown to originate from multiband chiral $p$-wave symmetry with weak anisotropy of pair amplitude.
The broad hump in conductance spectrum is a direct manifestation of the edge state peculiar to the chiral $p$-wave superconductivity.
\end{abstract}



Since the discovery of the ruthenate superconductor Sr$_2$RuO$_4$ (SRO)~\cite{Maeno1}, the symmetry of the pair potential has been a topic of hot debate.
The high anisotropy of transport properties indicates the quasi-two-dimensional electronic states of this compound.
Nuclear magnetic resonance~\cite{Ishida} and muon spin resonance experiments~\cite{Luke} suggest spin triplet pairing states with broken time-reversal symmetry. 
Taking into account the tetragonal crystal structure of SRO, the most plausible pairing state is chiral $p$-wave symmetry whose gap function is represented by $p_x \pm i p_y$~\cite{Maeno2}.
A number of experimental data that are consistent with the chiral $p$-wave symmetry have been presented.
However, the debate has not been settled because experimental data, such as the lack of spontaneous supercurrent~\cite{Kirtley}, challenge the interpretation in terms of the chiral $p$-wave symmetry.
\par
An important aspect of the chiral $p$-wave symmetry is that the phase of the pair potential evolves continuously with rotation in the $ab$-plane.
Thus, the chiral $p$-wave superconductor is a typical example of a topological superconductor.
One of the peculiar features of topological electronic states is the formation of the topologically protected edge states~\cite{Sato,Tanaka}.
In the cases of the chiral $p$-wave superconductors, the bulk states are fully gapped similarly to $s$-wave superconductors, whereas the formation of the gapless edge states is expected at surfaces perpendicular to the $ab$-plane (in-plane surfaces).
Revealing the edge states of topological superconductors is crucially important because various novel features, such as non-Abelian anyons and Majorana fermions, are expected to emerge at the edge.
\par
Tunneling spectroscopy is one of the most powerful tools for detecting the edge states of novel superconductors~\cite{Kashiwaya,Wei}.
However, experimental reports on 1.5K phase superconductivity in pure SRO are limited to several cases~\cite{Laube,Upward,Lupien,Suderow}.
The first reason is that high-quality superconducting SRO samples are synthesized not as thin films but as bulk crystals.
Therefore, conventional photolithography cannot be used for the fabrication of tunneling junctions.
The second reason is that the surface of the 1.5K phase is extremely fragile as revealed by 
scanning tunneling spectroscopy experiments~\cite{Pennec}.
This situation is unlike the 3K phase superconductivity in SRO-Ru eutectic crystals whose Ru inclusions have relatively inert surfaces.
Despite such difficulty, Refs.~\cite{Upward,Lupien} have presented the observation of a superconducting gap at the low-temperature-cleaved out-of-plane ($c$-axis) surfaces (Fig.~1(a)).
Moreover, fully gapped electronic states have been suggested from Al/SRO experiments of out-of-plane tunneling~\cite{Suderow}.
Laube, {\it et al.} reported the observation of conductance peaks using point contacts whose tunneling directions were not identified~\cite{Laube}.
On the other hand, no results detecting the edge states using well characterized in-plane tunneling junctions have been presented thus far.
Here, we report the fabrication process of in-plane tunneling junctions using SRO and verify the formation of the topologically protected edge states due to the chiral $p$-wave. 
\par
High-quality single crystals of SRO grown by a floating zone method were used to fabricate the tunneling junctions~\cite{Mao2}.
After the determination of crystal axes, the crystals were sliced into thin pellets with dimensions of about 1mm$\times$1mm$\times$20$\mu$m.
To achieve metal contacts between SRO and counter electrodes, the pellets were stored in a vacuum chamber, and crushed into small pieces with a hammer just before the Au electrode deposition.
Au films were subsequently deposited by DC sputtering on the crushed surfaces to protect the pristine.
The typical thickness of Au was about 400 nm at this stage.
The SRO pieces covered by Au were then removed from the chamber, and analyzed by laser microscopy and X-ray diffraction.
After fixing the selected pieces on SiO$_2$ substrates, the Au films were evaporated and four terminals were patterned using a focused ion beam (FIB).
A scanning ion microscopy image and a schematic of junction configuration are shown in Fig.~1(b).
Without using this $in$ $situ$ process, all the contacts between Au and SRO were non metallic.
In contrast, the contacts due to the present process showed almost temperature-independent metallic behavior (see Fig.~1(c)), and the onsets of superconductivity were discernible at 1.5K.
\begin{figure}[t]
\includegraphics[width=1\linewidth]{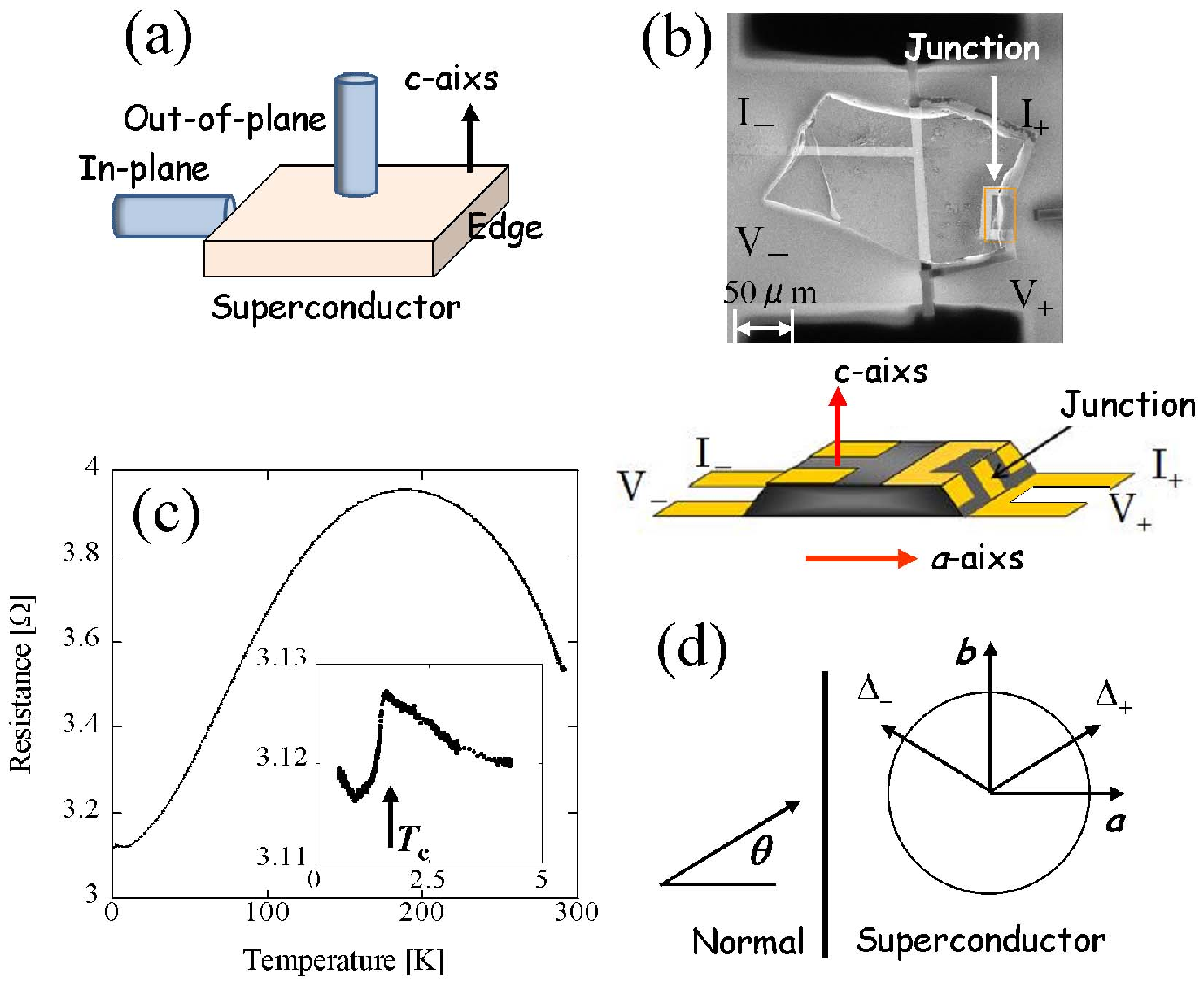}
\caption{\label{fig1}
(Color online)
(a) Schematic of in-plane and out-of-plane junctions.
(b) Scanning ion microscopy image ($300 \times 300$ $\mu$m$^2$) of a Sr$_2$RuO$_4$/Au junction fabricated on a SiO$_2$ substrate and a schematic of the junction structure.
The four terminal Au contacts are patterned by a focused ion beam, and the crystal axes are characterized by X-ray analysis.
(c) Temperature dependence of resistance of a Sr$_2$RuO$_4$/Au junction.
The inset shows the enlarged plot near $T_c$ ($\sim$1.5 K).
(d) Schematic of quasiparticle injection to an unconventional superconductor.
}
\end{figure}
\par
The pairing symmetries of the 3K and the 1.5K phases are suggested to be dissimilar to each other.
Unlike the bulk superconductivity of the 1.5K phase, the 3K phase is considered to be localized at the interface between SRO and Ru inclusions.
The appearances of a broad and a sharp zero-bias conductance peak have been reported for tunneling junctions fabricated on the 3K phase~\cite{Mao,Kawamura}.
Since these features are similar to those expected for the 1.5K phase, the effects of the 3K phase must be excluded carefully.
Figure 1(c) indicates the clear resistance change at 1.5K whereas no response at 3K.
Furthermore, Fig.~2 shows the magnetic field and temperature dependences of conductance spectra obtained on an SRO/Au junction.
The disappearances of the conductance peaks at $H_{c2}\sim 500G$ ($H$//$c$-axis) and $T_{c}\sim 1.5 K$ are consistent with the bulk properties of the 1.5K phase.
We detected the conductance change due to superconductivity in 5 out of 7 junctions.
At the interfaces of these junctions, the superconductivity in most of the areas is seriously suppressed probably owing to the high density of defects~\cite{Pennec}.
The survival of superconductivity at a restricted area contributes to the conductance channel of normal/insulator/superconductor (NIS).
\begin{figure}[t]
\includegraphics[width=1\linewidth]{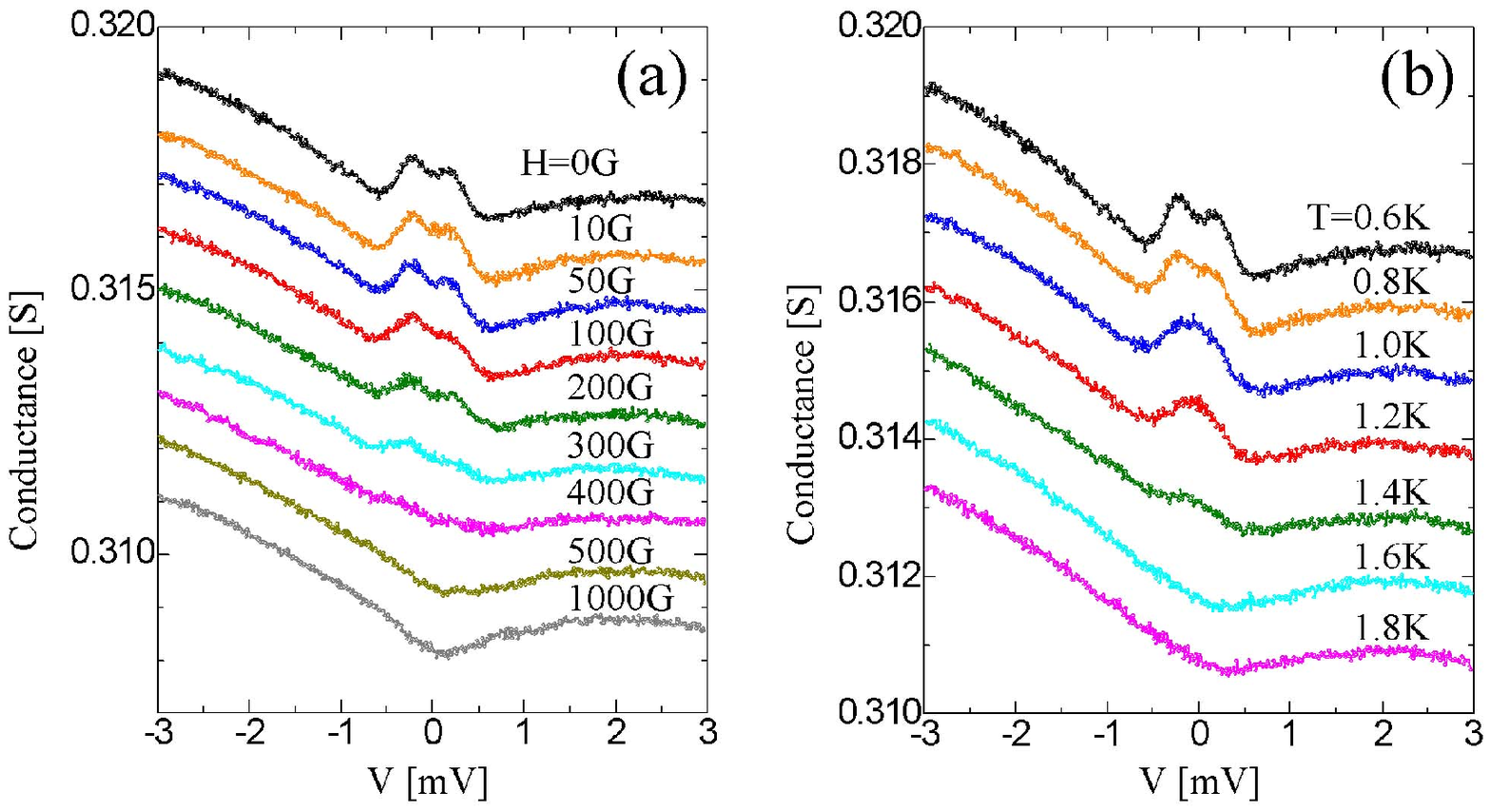}
\caption{\label{fig2}
(Color online)
Conductance spectra obtained on the same Sr$_2$RuO$_4$/Au junction as in fig.~1(c).
Each successive curve is vertically shifted by 0.001 for clarity.
(a) A series of spectra obtained in the magnetic field range between 0 G and 1000 G at 0.55 K.
(b) A series of spectra obtained in the temperature range between 0.6 K and 1.8 K in the absence of the magnetic field.
}
\end{figure}
\par
Figures 3(a)-3(c) show typical conductance spectra obtained on $a$-axis oriented junctions fabricated using the present process.
As a reference, the tunneling conductance spectrum of a (110)-oriented YBa$_2$Cu$_3$O$_{7-\delta}$ junction described in Ref.~\cite{Hkashiwaya} is shown in Fig.~3(d).
To clearly identify the response due to the superconductivity, the data are normalized by its background conductance in the normal states just above $T_{c}$.
The conductance spectra of SRO have common features: a broad hump spreading between $\pm$1 (horizontal axes are normalized in the figures), and small dips just outside the peak.
At the same time, the spectral shape near zero bias has sample dependence: (a) domelike peak, (b) split peak, (c) two-step peak.
Results (a) and (b) are reproducible because similar features have already been reported in the point contact spectroscopy~\cite{Laube}.
Although type (c) has not been presented in Ref.~\cite{Laube}, this feature is also reproduced on two of our junctions.
Therefore, we conclude that these three types are typical spectra intrinsic to the 1.5K phase.
\par
Next, we focus on the conductance spectrum of NIS formulated for a $\delta$-functional barrier model~\cite{Blonder}.
The barrier parameter $Z$ is represented by $mH/\hbar^2k_F$, where $m$, $H$, and $k_F$ are the electron mass, amplitude of the barrier potential, and Fermi wavelength, respectively.
The conductance spectrum is calculated using the reflection probabilities for the quasiparticle injection with energy $E$ (E=-$eV$, $V$: bias voltage) and incident angle $\theta$ with respect to the interface (Fig.~1(d)).
The conductance formula contains two distinct pair potentials $\Delta_{+}$ and $\Delta_{-}$, which correspond to the effective pair potentials for transmitted electron-like quasiparticles and hole-like quasiparticles, respectively~\cite{Kashiwaya,Yamashiro}.
The total conductance is given by integrating the angle-resolved conductance weighted by a tunneling probability distribution.
Except in the cases of low barrier limit, the peak corresponds to the energy level of the Andreev bound state (ABS) formed at the edge of the superconductor.
\begin{figure}[t]
\includegraphics[width=1\linewidth]{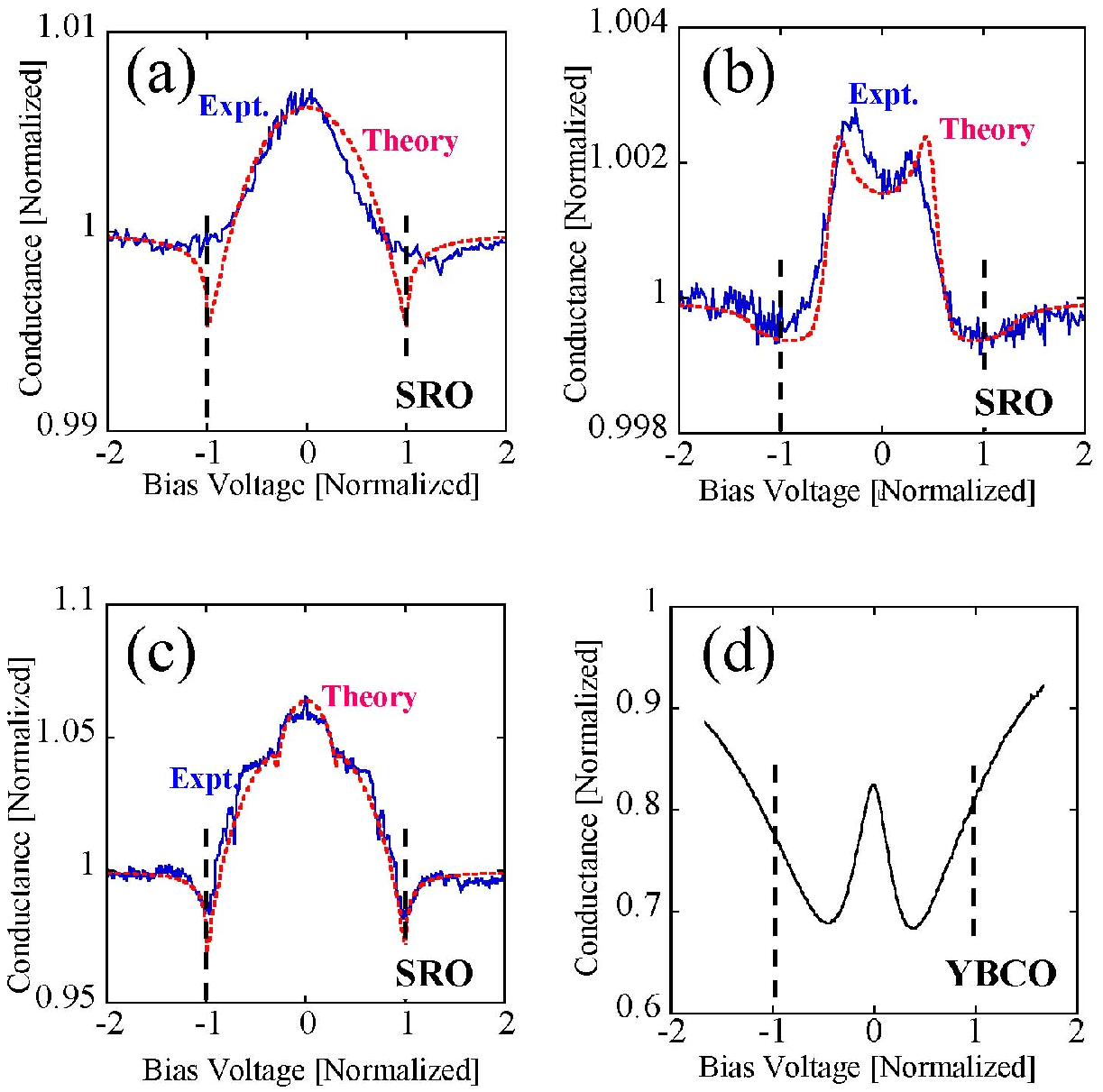}
\caption{\label{fig3}
(Color online)
Comparison between experimental conductance spectra obtained on Sr$_2$RuO$_4$ (SRO) junctions at 0.55K (solid lines) and theoretical spectra (dotted lines).
The vertical axes are normalized by their normal-state background values and the horizontal axes are normalized by (a, b) 0.7 mV and (c) 0.95 mV; the broken lines mark the gap amplitude as guides for the eye.
The dotted lines are calculated for (a) isotropic chiral $p$-wave symmetry with the barrier parameter Z=3 (domelike peak), (b) the chiral $p$-wave symmetry with a small gap amplitude modulation with Z=3 (split peak), and (c) two-band superconductivity with isotropic chiral $p$-wave symmetry with Z=2 (two-step peak).
(d) Experimental spectra obtained on an YBa$_2$Cu$_3$O$_{7-\delta}$ (YBCO) junction at 3 K cited from Ref.~\cite{Hkashiwaya}.
We normalize the vertical axis by the normal-state value and the horizontal axis by 18 mV.
}
\end{figure}
\par
To clarify the relationship between ABS and the pairing symmetry, we show in Figs.~4(a) and 4(b) the angle-resolved conductance spectra for the $d_{xy}$-wave (equivalent to $\pi$/4 tilted $d_{x^{2}-y^{2}}$-wave) and the chiral $p$-wave superconductors, respectively.
Fermi wavelengths of the normal and the superconductor are assumed to be equal for simplicity.
The conductance peak inside the gap region directly corresponds to the ABS formed between two pair potentials $\Delta_{+}$ and $\Delta_{-}$.
In the case of $d_{xy}$-wave superconductors, zero-energy bound states are formed independently of $\theta$, since the phase difference between the two pair potentials is $\pi$ ($\Delta_{\pm}=\pm \Delta_0$sin2$\theta$, where $\Delta_0$ is the amplitude of the pair potential).
Figure 4(c) shows calculated conductance spectra for $d_{xy}$-wave superconductors.
Despite the gap feature of the out-of-plane junction, the in-plane junction shows a sharp conductance peak at zero-bias level originating from the formation of the zero-energy ABS.
In contrast, the phase difference for the chiral $p$-wave superconductor is $\pi-2\theta$ ($\Delta_{\pm}=\pm \Delta_0 e^{\pm i\theta}$).
This feature leads to a formation of the diagonally distributed ABS as shown in Fig.~4(b).
The appearances of similar edge states are commonly expected for topological quantum states, such as superfluid $^3$He and Quantum Hall state.
These phenomena are interpreted as the consequence of ``bulk-edge correspondence'' peculiar to topological quantum states~\cite{Sato}.
Corresponding to the diagonal ABS formation, the conductance spectrum of in-plane junction shows a broad hump spreading all the way from -$\Delta$ to $\Delta$ despite the BCS-type conductance spectrum of the out-of-plane junction as shown in Fig.~4(d).
\begin{figure}[t]
\includegraphics[width=1\linewidth]{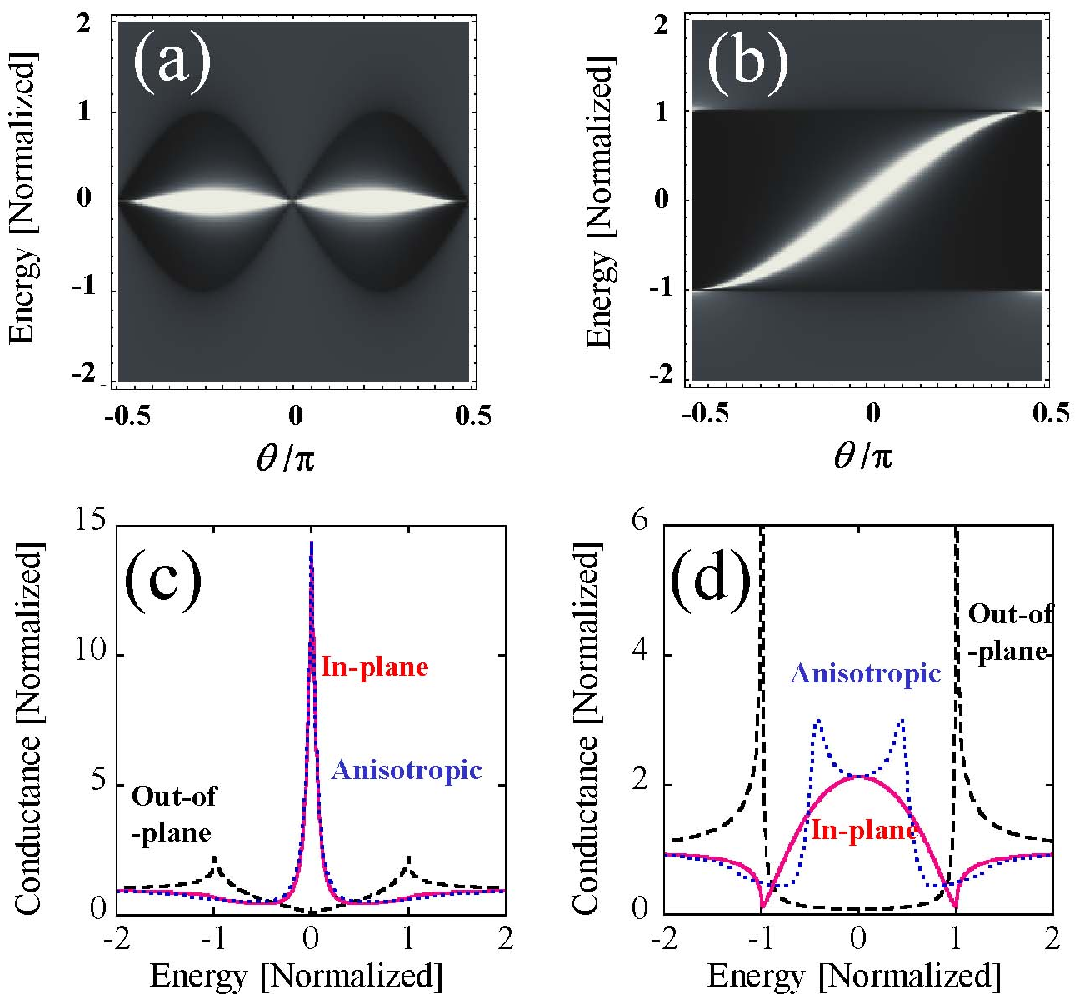}
\caption
{\label{fig4}
(Color online)
(a, b) Angle-resolved conductance spectra for (a) $d_{xy}$-wave superconductors and (b) chiral $p$-wave superconductors with the barrier parameter $Z$=4.
The vertical axis is the quasiparticle energy normalized by $\Delta_0$.
The brightness represents the magnitude of conductance of the junction.
Bright regions correspond to the conductance peak originating from Andreev bound states formed at the edge.
(c, d) Conductance spectra for (c) $d_{xy}$-wave superconductors and (d) chiral $p$-wave superconductors.
The solid line, broken line and dotted line represent the in-plane junction, out-of-plane junction, and in-plane junction with the anisotropic pair amplitude (see text), respectively.
}
\end{figure}
\par
The consistency between experimental data and theory is examined on the basis of the above discussion.
We can easily understand that the experimental conductance spectra showing a broad hump (Figs.~3(a)-3(c)) are consistent with the formation of the diagonal ABS in chiral $p$-wave symmetry.
In sharp contrast, the conductance spectrum of YBa$_2$Cu$_3$O$_{7-\delta}$ shown in Fig.~3(d) exhibits a narrow peak localized near zero-bias level.
Such behavior is apparently distinct from that of SRO, but consistent with the formation of $\theta$-independent zero-energy ABS of $d_{xy}$-wave symmetry.
\par
Next, we discuss the origin of the peak shape variety.
In the above, we assume an isotropic chiral $p$-wave superconductor with gap amplitude independent of the azimuthal angle in plane.
In reality, specific heat experiments~\cite{Deguchi} have clarified the presence of anisotropy in SRO. 
Since the effects of anisotropy on conductance spectra have already been discussed in Ref.~\cite{Sengupta}, here, we present an intuitive explanation based on a phenomenological model.
As the simplest model, we introduce the anisotropy by replacing $\Delta_0$ by $\Delta_0(1+0.3\cos4\theta)$ taking into account the tetragonal crystal structure.
The calculated spectra for $d_{xy}$-wave and $p$-wave symmetries are shown by dotted lines in Figs.~4(c) and 4(d).
Although the peak shape of the $d_{xy}$-wave symmetry is almost insensitive to the anisotropy, that for the chiral $p$-wave symmetry is seriously modified.
The origin of the difference is explained as follows.
In the case of the $d_{xy}$-wave symmetry, the phase difference is fixed to $\pi$ even in the presence of the anisotropy.
Thus, the ABS is fixed to the zero-bias level.
Contrastingly, since the phase difference can take an arbitrary value in the chiral $p$-wave symmetry, the energy level of the ABS is seriously modified owing to the amplitude variation.
This leads to the sensitivity of the peak shape on the anisotropy in the chiral $p$-wave symmetry.
In other words, the observation of various peak shapes in SRO is consistent with the chiral $p$-wave symmetry that has a rotating internal phase.
Actually, the calculated spectrum fits well to the experimental spectrum, as shown in Fig.~3(b).
\par
Finally, the origin of the two-step peak shape (Fig.~3(c)) is analyzed.
The Fermi surface of SRO consists of three bands labeled $\alpha$, $\beta$, and $\gamma$~\cite{Maeno2}.
The superconductivity is believed to originate mainly from the $\gamma$ band (active band), and induced superconductivity exists in $\alpha$ and $\beta$ bands (passive bands)~\cite{Agterburg}.
Thus the gap amplitude is larger for the active band than those in the passive bands.
We can incorporate the effect of the multiband to the conductance formula following the method presented in Ref.~\cite{Brinkman}.
The calculated result indicated by a broken line in Fig.~3(c) assumes the chiral $p$-wave superconductivity with two gap amplitudes ($\Delta_{active}$=0.93 mV, $\Delta_{passive}$=0.28 mV) and weighting factors of 0.8 and 0.2 for the active and the passive bands, respectively.
The ratio $\Delta_{passive} / \Delta_{active}$ obtained from the fitting is almost consistent with that in the specific heat measurements~\cite{Deguchi}.
The good agreement with the experimental data indicates that the origin of the two-step peak is reasonably ascribed to the multiband effect.
However, there remains a question as to why the peak corresponding to the passive bands is missing for spectra shown in Figs.~3(a) and 3(b).
A similar trend has been reported in point contacts of MgB$_2$; some of the experimental spectra show two gaps whereas the others show a single gap~\cite{MgB2}.
One possible explanation for such variety is the dependence of the weighting factor on the microscopic structure at the junction interface owing to the difference in the dispersion in each band.
Another possibility is the modification of the pair potentials owing to the presence of the edge~\cite{Yamashiro2}.
The pair potentials of the passive bands may be suppressed depending of the microscopic structure near the junction interface.
Details of these effects will be discussed in future work.
\par
In summary, we have successfully fabricated $a$-axis-oriented Sr$_2$RuO$_4$/Au junctions using an $in$ $situ$ sputtering process.
Conductance peaks detected in experiments are accounted for in terms of multiband chiral $p$-wave superconductivity with weak anisotropy of pair amplitude.
These results reveal the presence of the edge states due to the topological superconductivity of Sr$_2$RuO$_4$.
%
\par
We are very grateful to T. Nomura and A. A. Golubov for fruitful discussion.
This work was financially supported by a Grant-in-Aid for Scientific Research on Innovative Areas ``Topological Quantum Phenomena'' (No.~22103002) from MEXT, 
Grants-in-Aid for Scientific Research (No.~22340096 and No.~20221007) from JSPS, Japan.
%
%
%

%
%

\end{document}